# Stability analysis for nonlinear compressor system and active adaptive controller against surge with antisurge valve

Seyed Mohammad Hosseindokht*(mohammad.hosseindokht@upc.edu) -Department of Electrical Engineering, Universitat politecnica de Catalunya(UPC)

***Abstract* — In this paper, a compressor system is analyzed in order to show its characteristics and design a control scheme to improve its efficiency. A mathematical technique has been created to forecast the onset of surge and instability in a compressor chart, drawing from the nonlinear Greitzer and Moore model. This approach employs the phase plane and Jacobian matrix to identify both stable and unstable regions within the compressor, as well as to capture the limit cycle within the unstable region. A predictive analytical approach for anticipating compressor surge and instability is of great importance in system instrumentation and control. State space model is built up by nonlinear Greitzer equations. Validation from previous study about especial compressor will be considered for evaluation of mathematic method. Upstream flow acts as a disturbance to control loop and controller cannot satisfy desired requirements with flow variances, ergo it is essential that controller is adapted to new conditions. Since control signal is linearly related to system output, a PD controller is used to control compressor system. An adaptive PD controller is designed with MRAS method based on a reference model. Adaptive controller can stabilize compressor and increase its efficiency in the presence of any disturbances. Simulation results shows that an adaptive controller can provide good performance and convergence in case of speed changes by adapting gain parameters, and adaptive will be compared with normal PID. Finally, controller stability is investigated.**

## I. Introduction

Compressors are used in many industrial applications such as refineries [1], chemical and petroleum plants [2], and large scale refrigeration [3]. Compressor is very useful in energy management since it can inject determined air to fuel cell for producing electricity. Surge and stall are unfavorable phenomena in compressor that have adverse effects on the system performance [4]. Sometimes it is necessary to shut down and restart the compressor immediately to restore the operational level of the compressor and this could be very dangerous even infeasible in some cases like turbojets (air-plane) [5]. Reducing compressor flow increases its output pressure until maximum point in performance map is reached; from this point on more reduction in compressor air flow results in return flow due to more output pressure, which is called surge. Surge is an unstable phenomenon that happens when compressor input mass flow rate decreases or its output pressure increases at constant speed. High pressure in outlet and low flow inertia result in an intense reverse flow. Despite reverse flow, rotor transferred energy forces compressor to return to normal forward flow and higher pressure in outlet. However, by the same throttling condition, outlet pressure and reverse flow once again overcomes inlet flow. If the conditions that provoked the original surge event have not changed, the surge cycles will repeat [6]. This cycling process goes on very quickly and tends to cause drastic oscillations in the amplitude of some of thermodynamic and aerodynamic parameters [7]. Therefore, it is imperative to predict surge point, compressor performance map and unstable zone to design a controller that stabilize the process and improve its efficiency. Appropriate industrial instrumentation selection is based on the ability to estimate instability. Stability analysis of nonlinear system is based on the Greitzer state space model, Jacobian matrix and phase plane method [8] . Surge oscillations or limit cycle existence is also proved with Bendixson theorem [9–11]. Various control methods except adaptive method are used to deal with surge phenomenon. In a few works [12,13], meaning of adaption is not complete in comparison with MRAS (model reference adaptive system). There is a method of linearization for simplifying and compressor model and actuator are unknown and different with mentioned model and actuator in this paper. Moreover, there is not model analysis for understanding stable and unstable regions and surge phenomenon. These works [12,13] can not guarantee stability in deep and surge disturbances in unstable region. Previous controllers were fixed and could not adapt themselves with disturbances. It is clear that operating point changes with disturbances in compressor system, and since it is a nonlinear system, any changes in operating point can cause instability. Previous adaptive controls in the literature referred to compressor inlet adjustment based on desired operating point by some close-loop control system and did not use adaptive methods [14,15]. Some of previous works in the compressor field have used to cold and warm drivers to change controller type in order to adapt controller to changes in system states and called that adaptive controller [16]. In these works [17,18], compressor speed is used to control rotor speed by changing speed instead of flow and it was mentioned as an adaptive control. But, in this work, flow is measured as control variable and adaptive control [19] is considered. The compressor and the model that have used in [20] for compressor is different from the one that is used in here. In [20], a centrifugal compressor is illustrated while here an axial compressor is modelled. Therefore, proving stability and providing limit cycle information in the following compressor is of the contributions. In [20], there are no discussions about the control scheme. But here, we explain how to use the stability results in order to design an adaptive PD controller. In this work, mathematical methods such as Jacobian, phase plan and Bendixon theorem can investigate stable and unstable region of compressor. Validation data from previous study about especial compressor can clarify performance of mathematical method. An adaptive Proportional-Integral-Derivative (PID) controller is designed based on a reference model. Adaptive controller can stabilize compressor and increase its efficiency in the presence of any disturbance. Normal PID is designed for comparison with adaptive PID for evaluation of guaranteeing stability. Here, compressor model is not required since only inlet input is controlled. This simplifies the implementation and is an innovation in the control of a compressor system. Model Reference Adaptive Systems (MRAS) method based on MIT (Massachusetts Institute of Technology) rule [19] is the basis of adaption for controller design. The system stability is also proved. The main contributions of the paper are:

• Stability of the compressor is proved with Jacobian method.

• Limit cycle of the compressor is derived with Bendixson theorem.

• Validation data from previous study about especial compressor can clarify performance of mathematical method.

• An adaptive PD controller based on gradient descent method is designed and its stability is proved.

• Proving and analysing the effects of gain adaptation in convergence speed is an innovation in this work.

• Proving stability and analysing coefficients of adaptive PID based on averaging method is innovation. Moreover, comparison with normal PID can clarify performance of adaptive PID for guaranteeing stability.

The paper is organized as follows. In Section II, state space model of the system describing stable and unstable regions for nonlinear analysis is introduced. Section III Validates data from previous study about especial compressor. Section IV is dedicated to control design. It describes the reason for the linear PD controller choice, PD coefficients selection process and actuator and disturbance characteristics. Section V describes adaptive PD controller design. Section VI is simulated based on normal PID. Section VII includes concluding remarks.

## II. STABLE AND UNSTABLE REGIONS AND LIMIT CYCLE

Aerodynamic equations describing nonlinear state space model of the compressor are analyzed using Jacobean matrix and phase plane to determine the stable and instable domains of compressor and surge event. The compressor map $\psi_c(\varphi)$ is a potential function and it is a function of the compressor speed which relates mass flow rate ($\varphi$) to the pressure rise ($\psi_c$) in steady states conditions. The compressor goes through transient or dynamic condition to achieve steady states condition. It should be considered that $\psi_c$ and $\varphi$ are normalized functions of pressure rise and mass flow rate. There are especial performance map or steady states condition for one especial compressor in (1)[8,21].

$$\psi_c(\varphi) = 0.352 + 0.18[1 + 1.5(4\varphi - 1) - 0.5(4\varphi - 1)^3] \tag{1}$$

The relation between transient mass flow rate and pressure rise in nonlinear systems are described by a set of first order differential equations, introduced in references [21] and [8]. These equations are called Moore-Greitzer state space equations

$$\begin{aligned} f_1: &\quad \dot{\varphi} = 0.8(\psi_c(\varphi) - \psi) \\ f_2: &\quad \dot{\psi} = 1.25(\varphi - g\sqrt{\psi}) \end{aligned} \tag{2}$$

Where $\psi$ is plenum pressure and $g$ is a constant parameter and it depends on the throttle valve settings or compressor inlet mass flow rate (input). In the equations of the system, $g$ is not related to mass flow rate and pressure rise of plenum. In steady states condition, system's mass flow rate is equal to inlet mass flow rate [8]. In all equilibrium points, the following equations are valid

$$\psi_c(\varphi) = \psi \tag{3}$$

$$\varphi = g\sqrt{\psi} \tag{4}$$

Simplification of equations in equilibrium point leads to the following equation

$$\psi_c(\varphi) = \left(\frac{\varphi}{g}\right)^2 \tag{5}$$

Due to limited operating range of compressor ( $0 < \varphi < 0.8$ ), there is only one equilibrium point for the nonlinear system describing compressor (note that g is a positive constant). This indicates that Jacobian matrix method investigates global stability region of system; which means that for a stable system starting from any initial point, states of the system will converge to the only equilibrium point. If the system becomes unstable, starting from each initial operation point, states of the system will increase and they will become infinite. In other words, stability of this system is investigated globally. It should be noted that if there was more than one equilibrium point for the system, analysis of the stability would be local and would only be suitable for some operation point. In this condition, Jacobian matrix is not enough for analysis; hence, phase plane method should be applied to describe global stability of the system. To calculate Jacobian matrix, partial derivatives are used:

$$J = \begin{bmatrix} \frac{\partial f_1}{\partial \varphi} & \frac{\partial f_1}{\partial \psi} \\ \frac{\partial f_2}{\partial \varphi} & \frac{\partial f_2}{\partial \psi} \end{bmatrix} = \begin{bmatrix} 0.864 - 0.864\left(4\varphi - 1\right)^2 & -0.8 \\ 1.25 & 1.25\ (-g\frac{1}{2\sqrt{\psi}}) \end{bmatrix} \tag{6}$$

Eigenvalues of the Jacobian matrix are the roots of the quadratic equation:

$$Det\ (sI - J) = 0 \tag{7}$$

$$\rightarrow \begin{aligned} & s^2 + s[0.625\frac{g}{\sqrt{\psi}} - 0.864 + 0.864\ (4\varphi - 1)^2] \\ & +\ 0.625\frac{g}{\sqrt{\psi}}[-0.864 + 0.864\ (4\varphi - 1)^2] + 1 = 0 \end{aligned} \tag{8}$$

Where $I$ is a 2-by-2 identity matrix. Establishment of the equilibrium in equations (3) and (4) obtained from substitution of unknown parameters $\psi$ and $g$ in equation (8) results in equation (9) which only depends on $\varphi$.

$$\begin{aligned} & s^2 + s\ [0.625\frac{\varphi}{\psi_c(\varphi)} - 0.864 + 0.864(4\varphi - 1)^2] + \\ & 0.625\frac{\varphi}{\psi_c(\varphi)}\ [-0.864 + 0.864(4\varphi - 1)^2] + 1 = 0 \end{aligned} \tag{9}$$

The discriminant of the above quadratic equation ($\Delta$) that is calculated in Eq. (10), has negative value in the compressor's working range of mass flow rate ($0 < \varphi < 0.8$) as shown in Fig. 1.

$$\begin{aligned} & \Delta = 0.39(\frac{\varphi}{\psi_c(\varphi)})^2 + \left[-0.864 + 0.864(4\varphi - 1)^2\right] * \\ & \left[-0.864 + 0.864(4\varphi - 1)^2 - 1.25\frac{\varphi}{\psi_c(\varphi)}\right] - 4 \end{aligned} \tag{10}$$

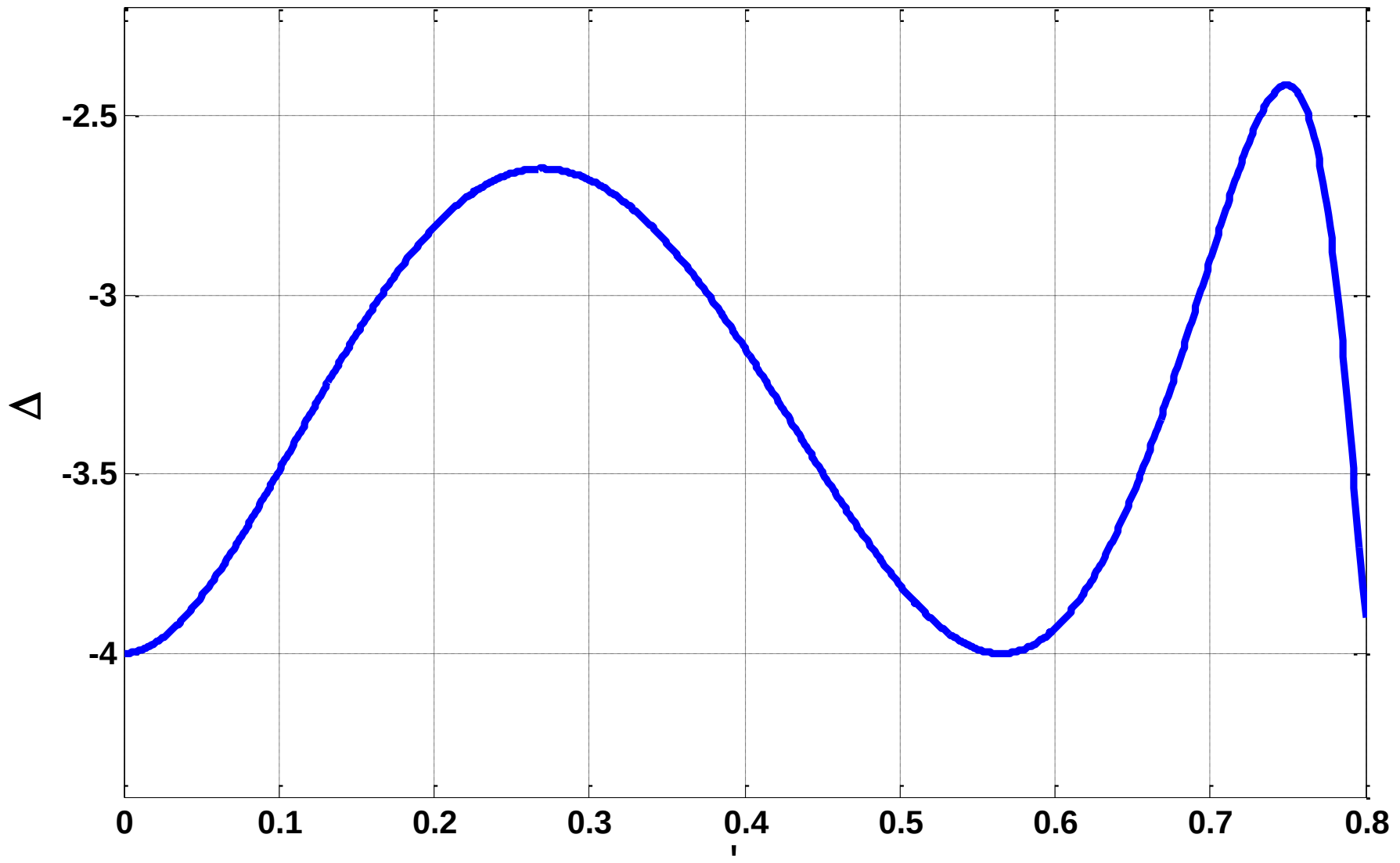

**Fig. 1.** ***Variations of Δ versus mass flow rate***

Fig. 1 shows the variation of discriminant in different mass flow rates. Due to the negative value of (10), the eigenvalues of the system are complex numbers. The real part of the eigenvalue varies by mass flow rate

$$\text{Real Part} = -[0.625\frac{\varphi}{\psi_c(\varphi)} - 0.864 + 0.864(4\varphi - 1)^2] \qquad (11)$$

The sign of the real part of the eigenvalue determines the stability of the operating point. Variation of the real part of the eigenvalue by mass flow rate according to equation (11) is demonstrated in Fig. 2. Negative sign of the real part of the eigenvalue indicates that the working point is a stable focus one.

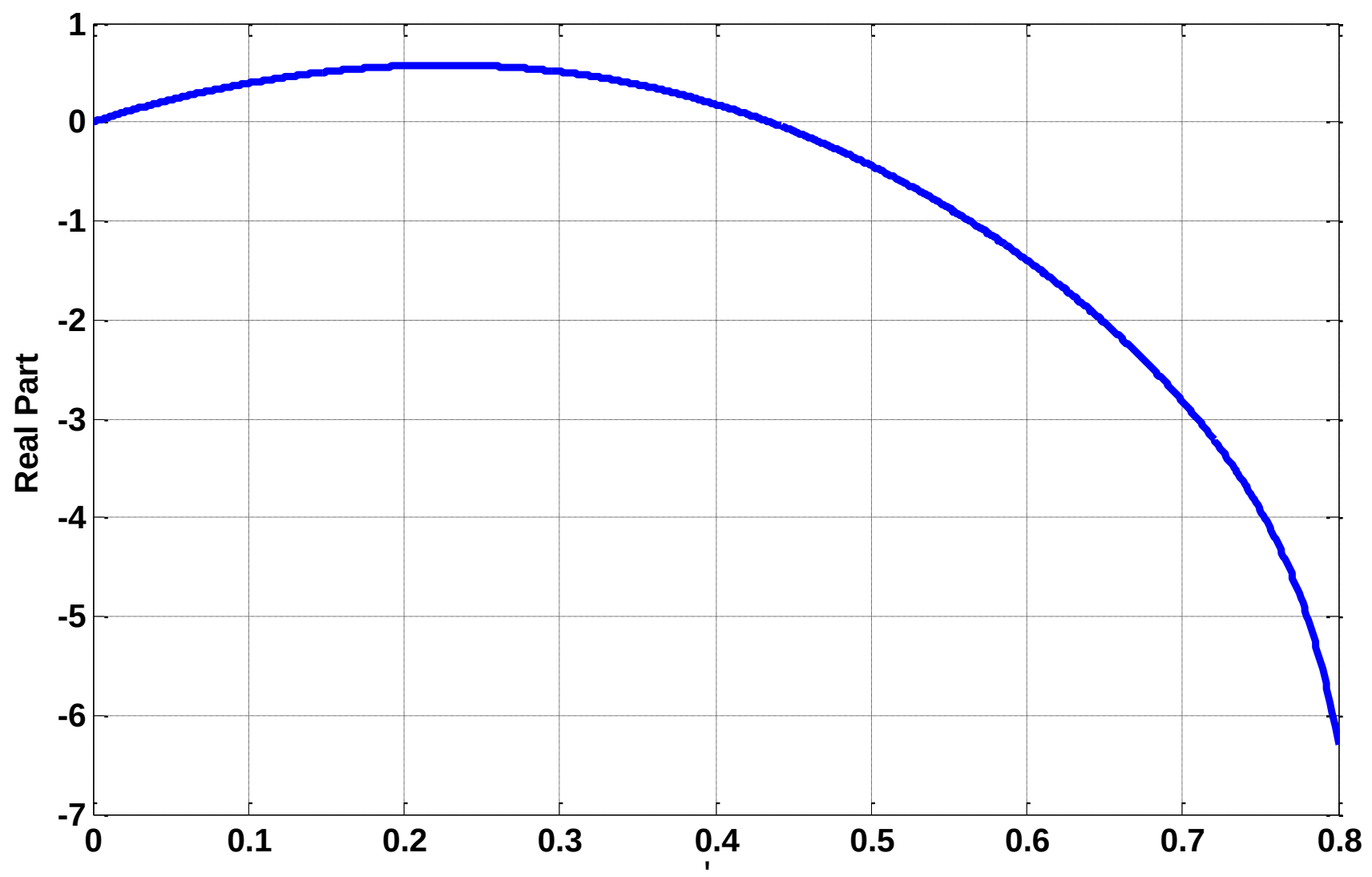

**Fig. 2. Variations of real part of eigenvalue versus mass flow rate**

Based on the calculated results depicted in Fig. 2, if non-dimensional inlet mass flow rate is more than 0.43, the compressor performance is completely stable and far from surge point and the status of compressor's working point on characteristic curve (performance map) remains stable. In other words, starting from any initial point with different mass flow, the system tends to a new stable state (a new steady states condition on characteristic curve) by changing working point of the compressor and its stability is guaranteed. It is obvious that Jacobian matrix method proves stability of any system locally, but according to the state space model (Greitzer equations), g is a function of throttle settings and in fact it is a function of states in equilibrium point (see Eq. (3) and Eq. (4)). In this case, equilibrium point is not stationary and it depends on the system state ($\varphi$). Eigenvalue of the Jacobian matrix is also not constant and is a function of $\varphi$. Therefore, its favorable sign with respect to all variations in mass flow rate was demonstrated so stability in all operating ranges of the compressor is guaranteed.

Global stability of the system is investigated by phase plane method too. The system's dynamics indicate that surge and instability start at about $\varphi = 0.43$. As a result, for the amounts of $\varphi$ less than 0.43, the working point lies to left side of surge line. In this condition the mass flow rate and pressure rise values oscillates and amplitude of oscillation increases. For example, on the unstable point with $\varphi = 0.4$ and $\psi = 0.6746$ (on left side of maximum pressure rise) variations of mass flow rate, pressure rise and the amplitude of their oscillation increases. This event is shown in Fig. 3 and 4.

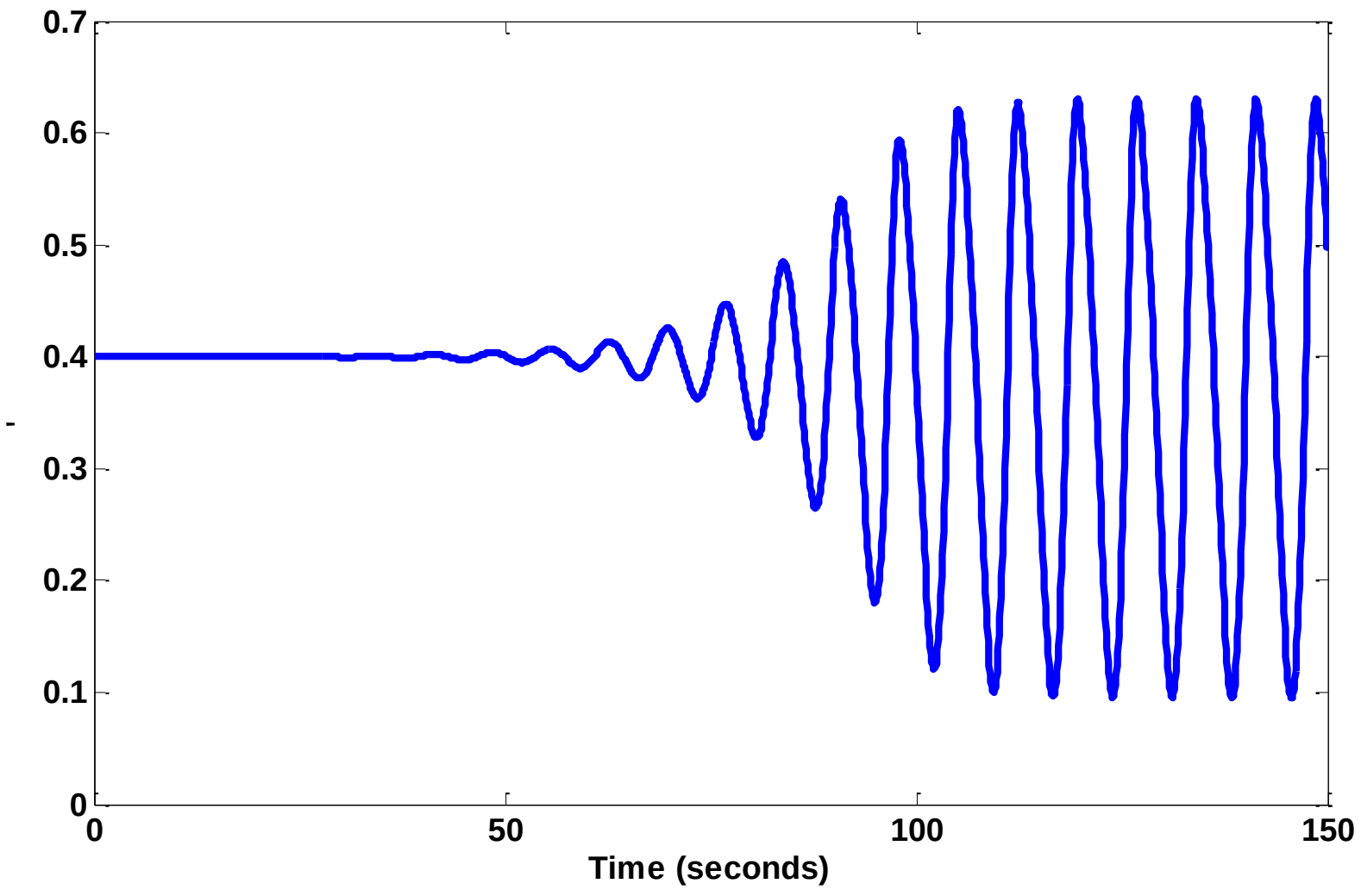

**Fig. 3. Flow variations within unstable region**

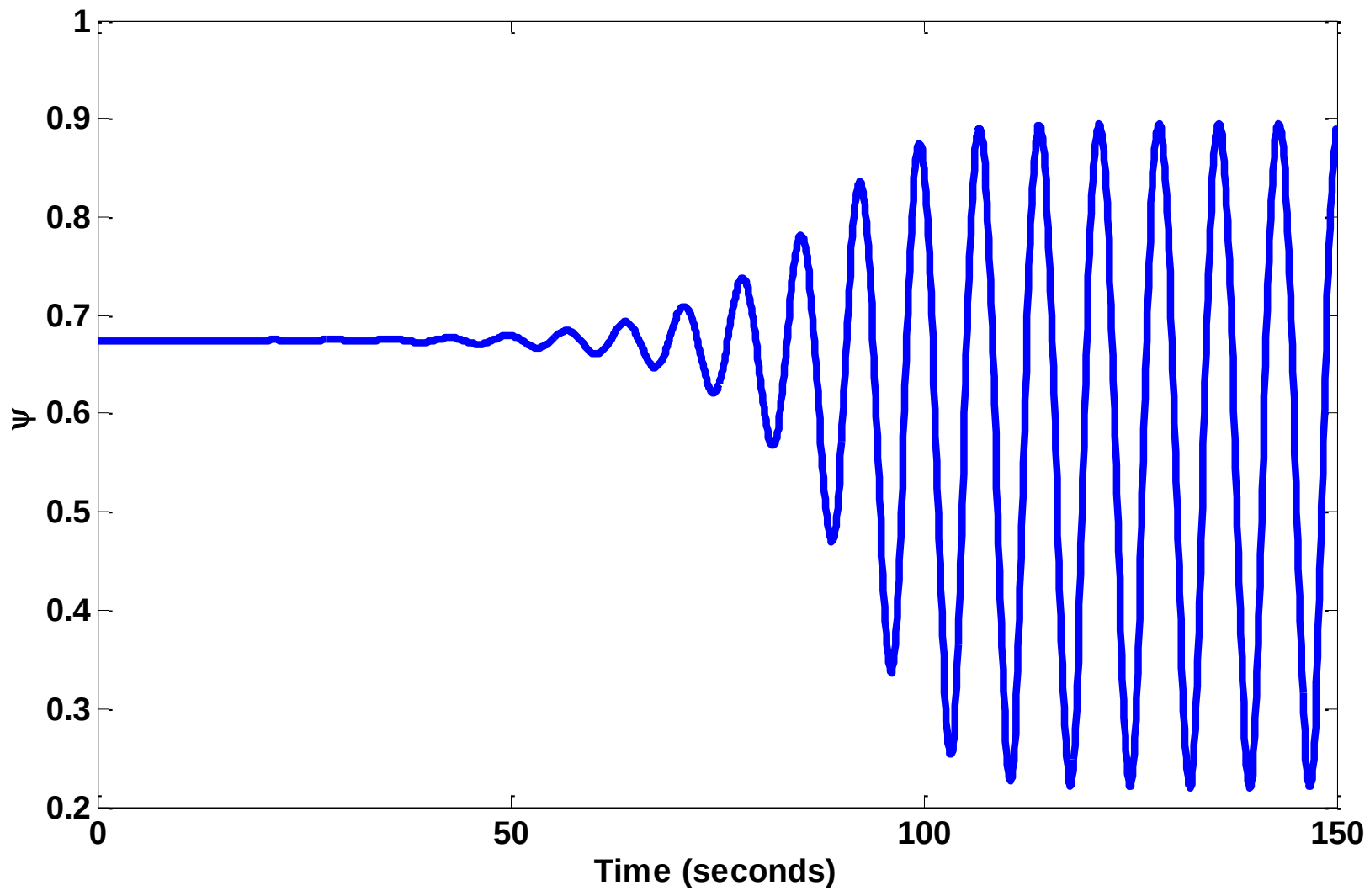

**Fig. 4. Pressure variations within unstable region**

Existence of limit cycle is important to investigate possibility of increase in amplitude of oscillations based on Bendixson theorem [22]. For this purpose, equations (4) and (5) are used to extract the following relation.

$$R = \frac{\partial f_1}{\partial \varphi} + \frac{\partial f_2}{\partial \psi} = 0.864 - 0.864\left(4\varphi - 1\right)^2 + 1.25(-g\frac{1}{2\sqrt{\psi}}) \\ = 0.864 - 0.864\left(4\varphi - 1\right)^2 + 1.25(-g^2\frac{1}{2\varphi}) \qquad (12)$$

Equilibrium state (Eq. (3) and (4)) is used to derive equation (12), therefore it is only a function of $\varphi$.

Amplitude enhancement of oscillations based on fluid mechanics and principles of aerodynamics is possible. As shown in Fig. 5, if compressor works in stable zone ($\varphi > 0.43$) sign of equation (12) is negative and it does not change. Consequently, according to Bendixson theorem [22], limit cycle does not exist in stable zone. However, in lower mass flows and in unstable zone, sign of equation (12) changes and limit cycle may emerge.

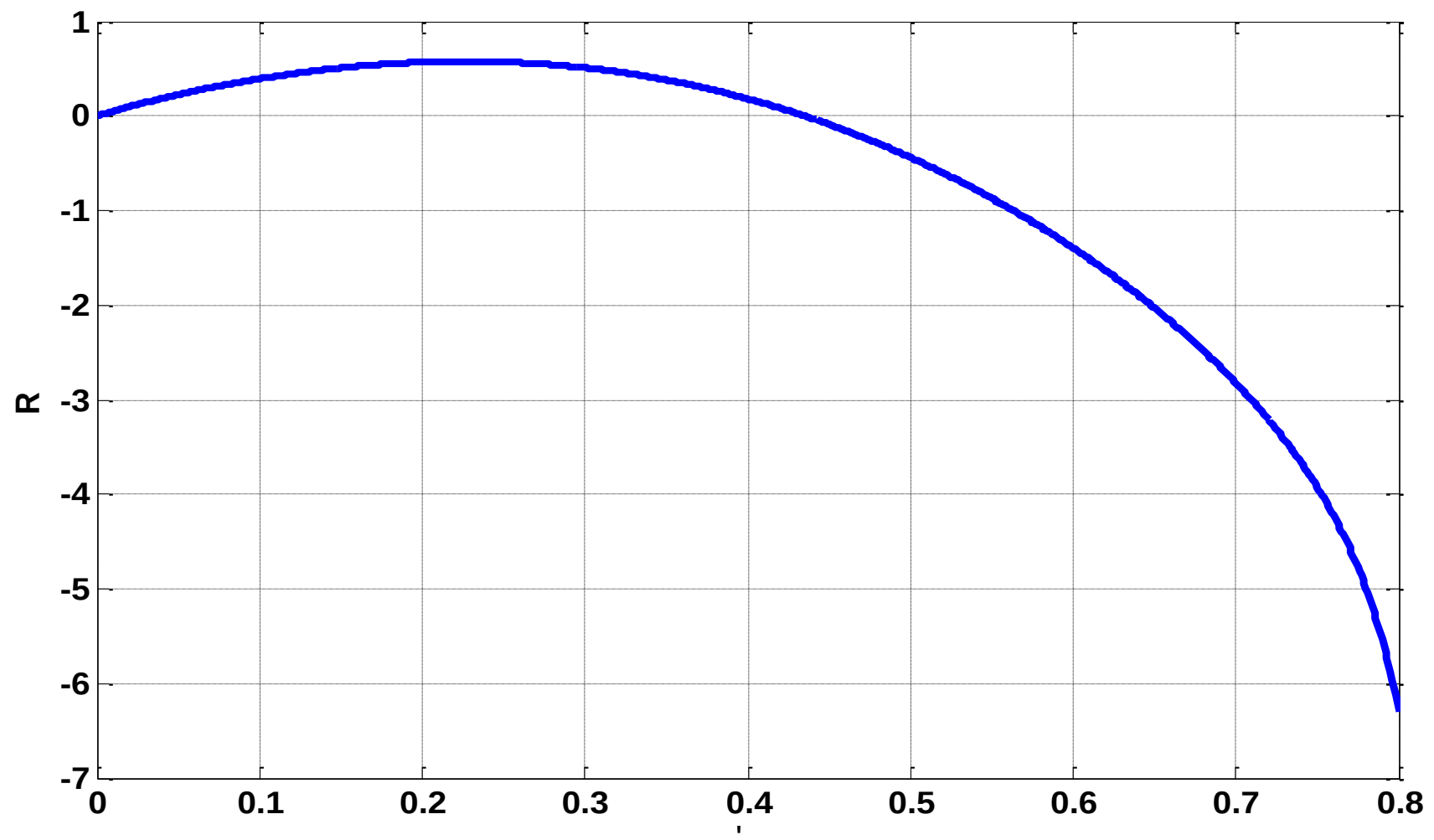

**Fig. 5. Variations of R factor versus flow**

Phase diagram plot indicates existence of limit cycle in unstable zone. If operating point of the compressor or equilibrium point tends to the unstable zone, mass flow rate and pressure rise variations will oscillate. Despite the fact that the mass flow rate is in unstable zone, operating points which is an unstable focus one, turn into limit cycle and amplitude of the oscillations remains constant. Therefore, due to unstable focus inside limit cycle, the cycle remains stable. Existence of Limit cycle is proved with phase plane diagram in Fig. 6.

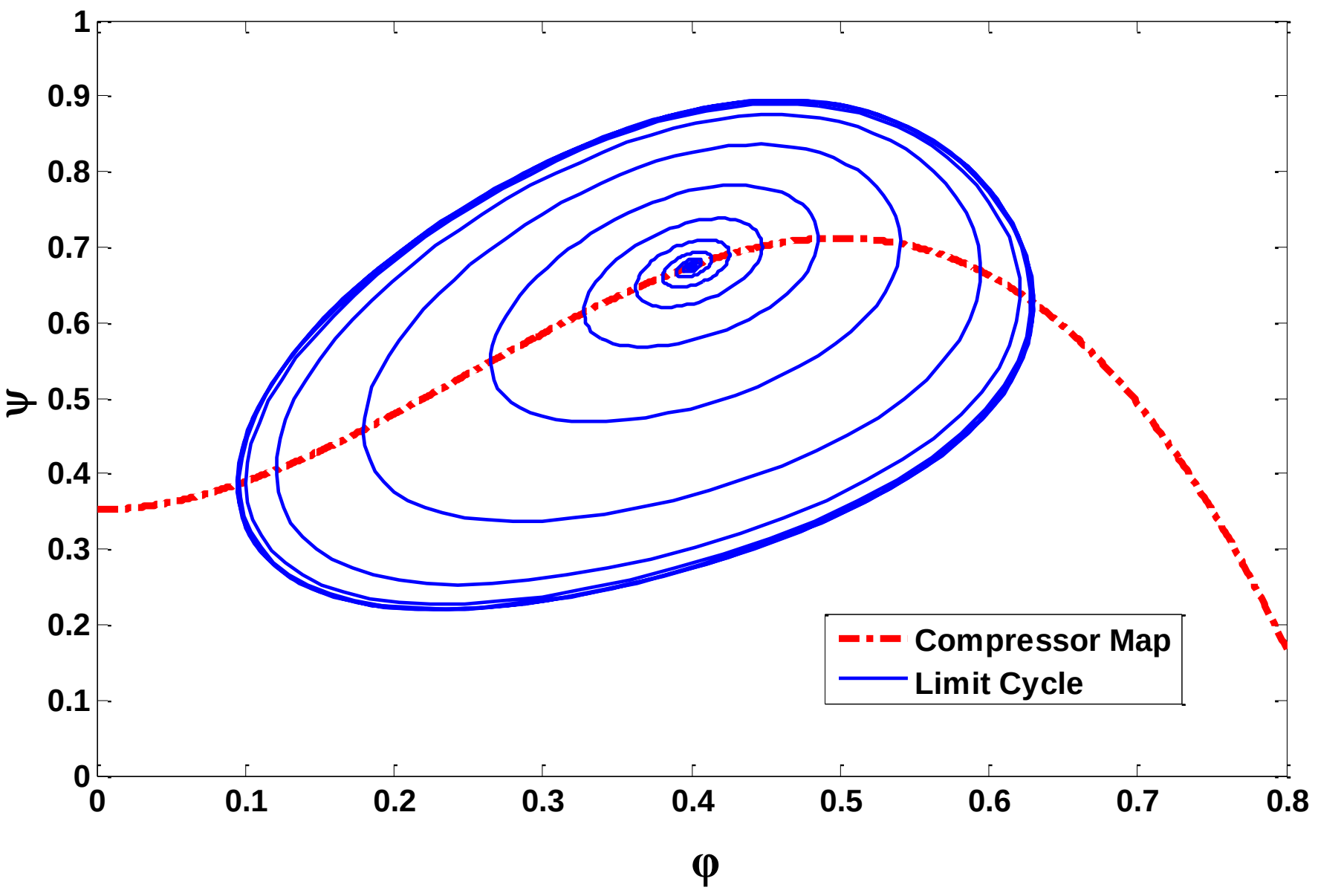

**Fig. 6. Limit cycle within unstable region**

## III. *VALIDATION OF* THE MATHEMATICAL METHOD

In order to validate the mathematical method, the mass flow which displays surge inception is compared to what is obtained through experiment by Koff. The case study is a three-stage compressor that for more detail, one can refer to [23]. Koff by using an unstalled method provides a performance map along with anticipated stall point and the axisymmetric peak point. Koff validation method showed that the compressor unstalled performance is for $\Phi$ >0.48. In fact, according to his obtained data, the inception of instabilities is for amounts of "$\Phi$" that are less than 0.48. The related characteristic is shown in Fig7. In the present study, despite many simplifying assumptions, the instability inception is about $\Phi$<0.43. The lower mass flow coefficient which is surge inception is due to many reasons. Firstly, in the utilized mathematical method (Greitzer model), surge and stall inception are considered simultaneously. Secondly, as shown in Fig7, there are two concepts of stall point. One concept is a true stall point which is a function of compressor geometry only.

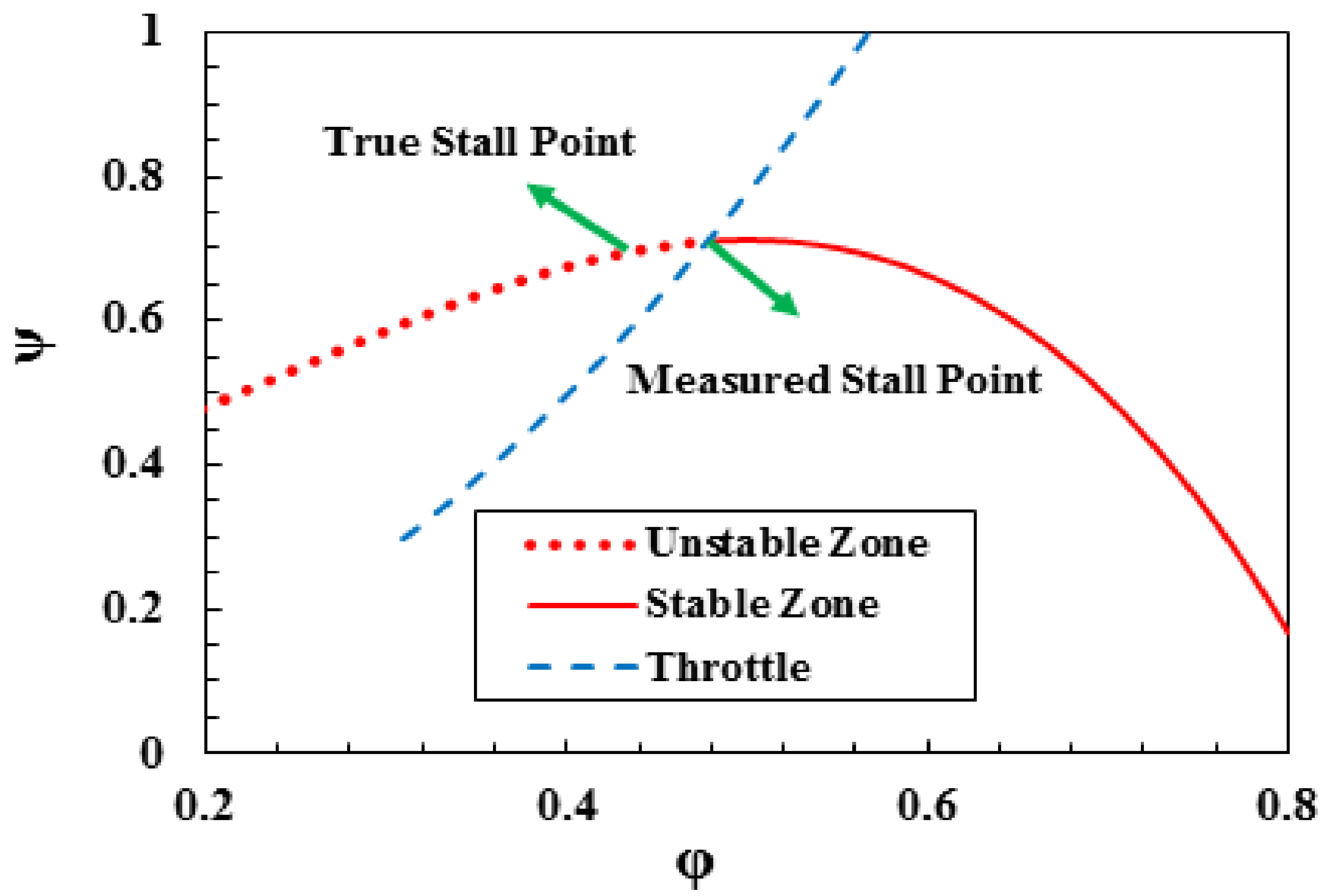

**Figure7. True and measured stall point**

The other concept of stall point is called measured stall point which exists on a positively sloped portion of characteristic [23]. Since the compression system instability may occur first and then include compressor, these two points are not a coincidence. However, the true stall point occurs in less mass flow coefficient, because it is a function of compressor geometry only. In other words, the true stall point remains hidden from the compressor test. It means that the true stall point is not obtained in the experiment [23]. Moreover, as shown in the characteristic of Fig7, others suggest that stall can occur at the peak pressure coefficient because the system is dynamically unstable [24]. It was mentioned in [23] that stall is mostly caused by wall stall which is not considered in two-dimensional criteria of maximum pressure coefficient. Consequently, this criterion for the inception of instabilities is a simple moderate prediction.

## IV. PARTS OF CONTROL LOOP

### *A. Control Variable*

System output measured by sensors is controlled variable. Compressor map maximum point is chosen to be operating point for the best efficiency. Equilibrium points on the compressor map are also used to determine stability as proved in section II. Both pressure rise and mass flow rate are dependent variables so it makes no difference which one is the controlled variable. Given the compressor dynamic in two variable state space, its behavior is oscillatory and pressure and flow oscillate to reach steady state for all initial conditions. For example, in Fig. 7 ,with initial condition $[\varphi_0, \psi_0] = [0.63, 0.62]$, nonlinear system reaches steady state in $[\varphi_{ss}, \psi_{ss}] = [0.51, 0.71]$.

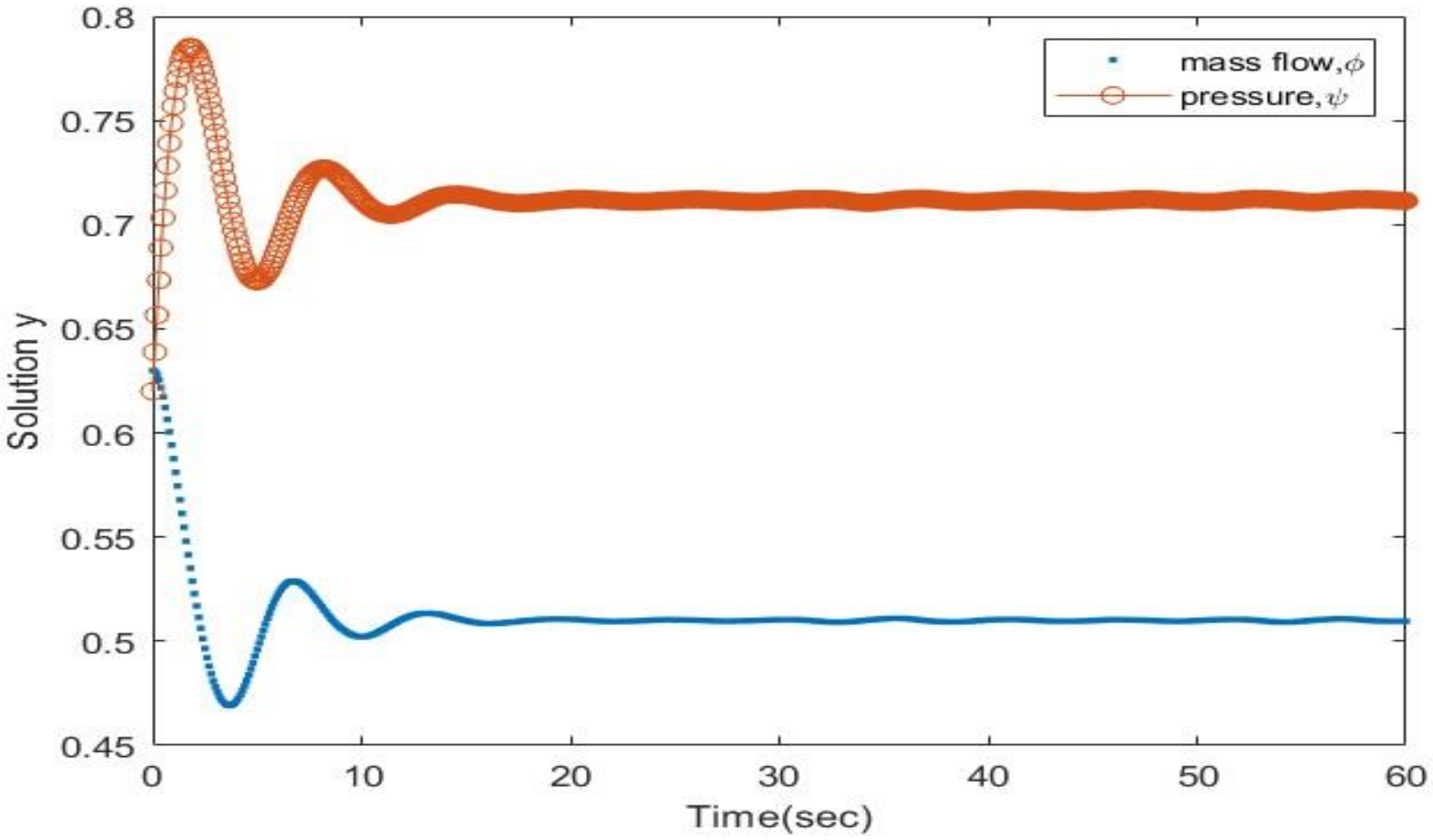

**Fig. 7. Oscillatory behavior of states in transient mode**

If the states are considered controlled variable and sensors measure them, oscillatory behavior of the variables cannot exactly be used to determine operating point characteristics. Input mass flow rate has stable behavior, if input mass flow rate changes then operating point moves on the map and there won't be an oscillatory behavior for the mass flow rate. The basic difference in input and output mass flow rate is important point for input mass flow rate so it can be chosen as control variable. If the input of the compressor is measured and measurements of the input mass flow rate is fed to the controller, compressor and its relations can be ignored in control system design. The only assumption is that the compressor is working in stable zone ($\varphi > 0.43$). However, if by any circumstances in a short period of time, $\varphi$ becomes less than 0.43, the designed controller is able to take the system back to the stable region.

If the pressure rise is chosen as control variable, compressor relations cannot be ignored because the output pressure is required in equations. If the input mass flow rate is selected as controlled variable then oscillation errors of the output mass flow rate (state of the system) and compressor is going to be eliminated and the system's heavy nonlinear state space equations vanishes; hence, equations become simple enough for a straight forward controller design. Moreover, when compressor enters to surge if output mass flow is measured as control variable, there are some oscillations(before reaching steady state) more than surge or unstable region, and output mass flow shows operation of compressor in stable zone and this is wrong and can cause bad decision for avoiding surge. Then this is better input mass flow is considered as control variable, because this is input (throttle parameter) and there is steady state mode for this without dynamic and oscillations of states in Grietzer equations. Totally, input mass flow(steady state) and output mass flow(plenum mass flow) are equal after transient or dynamical mode. This transient mode can cause oscillations in output mass flow that can't show correct mass flow. Then choosing input mass flow is better for sensing mass flow of compressor for controlling this.

### *B. Actuator, Disturbance and Reference*

Control system actuator is an anti-surge valve that returns output mass flow rate to the compressor input and it moves operating point to stable zone [25,26]. Surrounding or upstream mass flow rate of the compressor is a disturbance signal. Both disturbance and returned mass flow rate from anti-surge, accumulate and enters as a single input to the compressor. That's why anti-surge accumulates with disturbance and doesn't reduce disturbance. Hence, actuator acts one-sided and puts limitation on the control signal. Not all operating points can be sustained with non-zero steady state error. Valve operates in a manner that is never actually closed and its value is always more than zero. Actuator malfunction is undetected if it is closed or there is no feedback signal.

Minimum and maximum mass flow rate of the valve is 0.05 and 0.25, respectively. Most of the industrial valves are modeled using delay and time constant. In this work time constant is supposed to be 2 seconds and there is no delay with a transfer function of [25,26]

$$G_v(s) = \frac{1}{2s+1} \tag{13}$$

Actuator is limited to maximum and minimum operating value; therefore, saturation makes it a nonlinear element. Maximum efficiency is 0.5 but reference or desired point is chosen to be 0.55 with a safe margin from maximum value which is instability boundary. The system will encounter no problem if sensor, actuator and controller operate improperly or slowly.

### *C. PID controller*

Input mass flow rate is controlled variable or output of control loop and anti-surge mass flow rate is manipulated variable or command signal to the compressor. The compressor input (controlled variable) is summation of disturbance and command signal, which means that the relation between output of the control loop (control variable) and command is linear. Therefore, if summation of actuator and compressor input is considered as process, the relation of process input (control signal) and its output (compressor input) is linear. Hence, a linear PID controller is the best choice among other controllers.

$$G_{PID}(s) = \frac{u(t)}{e(t)} = K_p + K_i \frac{1}{s} + K_d s = K_p \left(1 + \frac{1}{T_i s} + T_d s\right) \tag{14}$$

's' is Laplace domain. Where $e(t)$ is the error signal that fed to the controller and $u(t)$ is the control signal that is produced by controller and has to be given as input to the plant. PID coefficients ($K_p, K_i$ and $K_d$ or $K_p, T_i$ and $T_d$) can be determined by considering different design criteria such as stability, response time, steady state error, overshoot, optimization, etc. As much as $K_p$ is increased, valve is going to be opened more and plant stabilizes in surge state faster, but more energy consumed. Derivative part provides best estimate of the future trends of the error signal $e(t)$, based on its current rate of change. Therefore, by adding term $K_d$ to the controller, the system will work faster. Actuator can't follow all of the PID command signals, for example it can't produce negative mass flow rates. If operating point is more than reference value, desired point is not accessible with minimum mass flow of anti-surge, unless negative command is introduced or valve offloads the input mass flow rate which is not feasible. Any increase in $K_i$, results in summation of previous effects and present errors and it results in a later sequential command and therefore, controller speed reduces. That's why $K_i$ is chosen to be zero and PID is converted to PD. Disturbance variations is effective in control, as compressor input decreases rapidly, system enters unstable region faster and controller must open the valve more and more so that the system becomes stable. In a compressor, deviation from set point is more important than variations in error itself, so error deviation effects should be less than effects of error itself and derivation gain should be smaller than proportional gain. Especially when operating point oscillates about set point and stability and efficiency is good, valve operation is not essential. It saves energy and valve's life will increase.

This work is based on stabilization, hence the controller gains can be determined by first Ziegler-Nichols (Z-N) method [27]. In this method, at first the response of open loop system $G_{valve}(s)$ is derived (Fig. 8). Then, two constants (L and T) are determined by drawing a tangent line at the inflection point of the step response curve and finding the intersections of the that line with the time axis and the steady-state level line. Then, we can use Z-N rule table as shown in Table 1 to determine PID coefficients.

**Table 1. Z-N First method Tuning rule**

| | $K_p$ | $T_i$ | $T_d$ |
|---|---|---|---|
| **P** | $\frac{T}{L}$ | $\infty$ | 0 |
| **PI** | $0.9\frac{T}{L}$ | $\frac{L}{0.3}$ | 0 |
| **PID** | $1.2\frac{T}{L}$ | $2L$ | $0.5L$ |

These initial parameters will typically provide a response with an overshoot of the order 25% with a good settling time. We may then start fine-tuning the controller using the rules that relate each parameter to the response characteristics, as discussed earlier. Note that this method has no direct relation for PD coefficients. Accordingly, PID coefficients were calculated by Z-N method, then set $K_i$ to zero. This modification increases response time so the differentiator gain $K_d$, can be reduced to compensate the unfavorable effect.

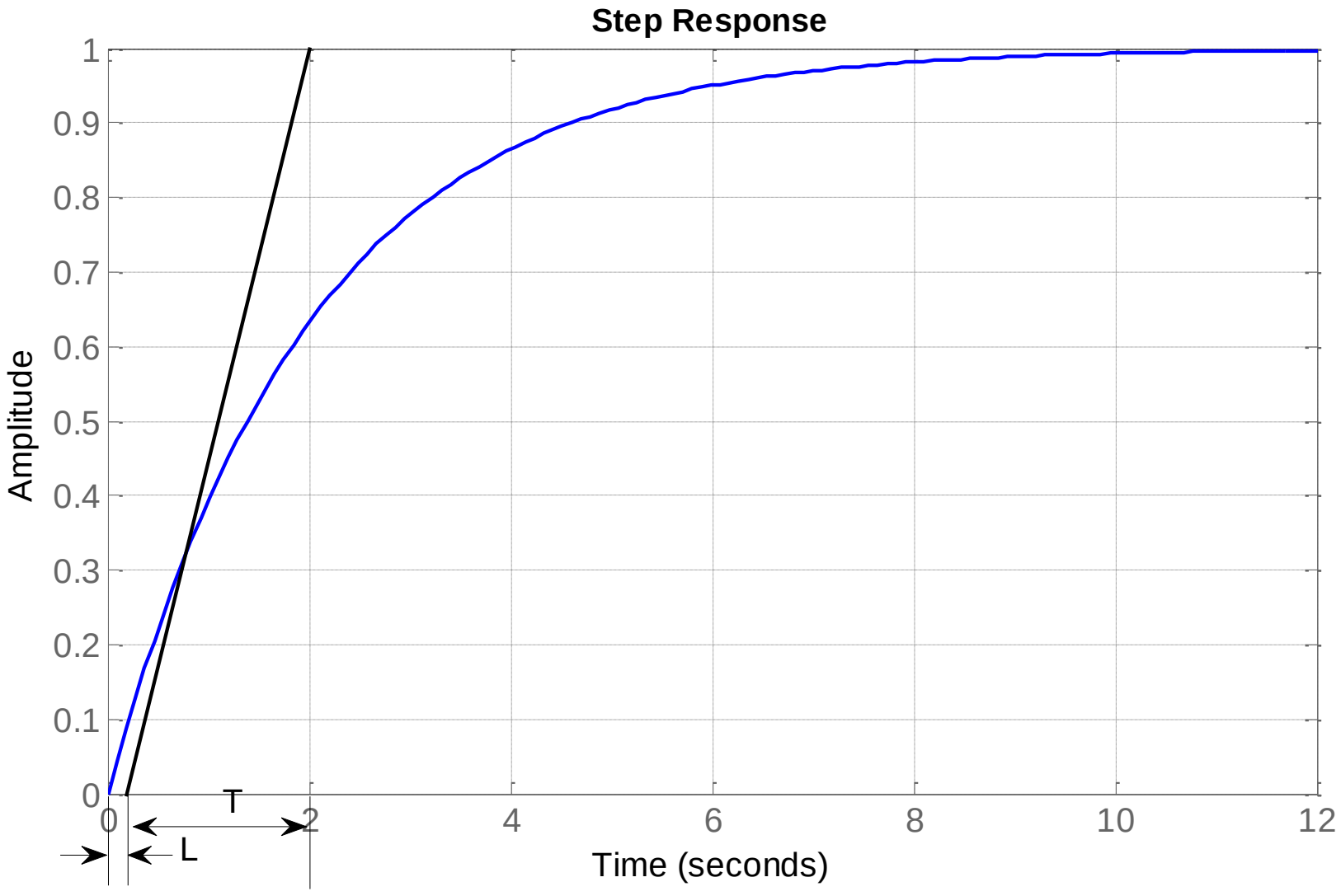

**Fig. 8. Deriving step response for designing PID controller using Z-N method**

$$T = 1.79, \quad L = 0.213 \;\rightarrow\; K_p = 1.2\frac{T}{L} = 10, \quad K_d = K_p T_d = 0.5 K_p L = 1 \qquad (15)$$

Gain choices of $K_p = 10$ and $K_d = 0.7$ leads to system response with rise time $t_r = 0.5\,(\text{sec})$, settling time $t_s = 0.9\,(\text{sec})$ and steady state error $e_{ss} = 0.09$. It is necessary to have a large ratio of derivation to proportional gain, the choice of $K_p / K_d = 15$ is satisfactory.

## V. Adaptive Control

### A. Designing Adaptive Controller

Disturbances change compressor input; consequently, operating point changes. In this case, fixed controller designed in section III with its features like $t_r$ and $e_{ss}$ may not be able to guarantee stability and efficiency. It means that controller has a low speed and this is dangerous because it cannot stabilize the system. Increasing system stabilization speed results in reducing steady state error.

Increasing $K_p$ decreases steady state error and increases response time but it is an expensive choice. Controller design would satisfy acceptable criteria including optimization, speed, steady state error, stability and efficiency. Model Reference Adaptive Systems (MRAS) method based on MIT (Massachusetts Institute of Technology) rule [16] is the basis of adoption for controller design. A schematic diagram of the proposed controller is shown in Fig. 9. First, a Reference model must be selected. To have settling time, rise time and steady state error as stated in III, the transfer function of the reference model is selected as

$$G_m = \frac{25}{s^2 + 8.5s + 25} \qquad (16)$$

Gradient descend method or MIT rule adds to the degrees of freedom for controller design, therefore it is a better adaptation law for increasing coefficients of the linear controller in the presence of a nonlinear actuator and it improves the convergence to the reference model. When system is nonlinear, as parameters (coefficients) of linear controller increases, gradient descend method performs better in adaptation of the system and it converges faster to a reference model. PD controller has maximum 3 parameters. PD controller structure with three parameters of $k_1$, $k_2$ and $k_3$ is as follows

$$u = k_1 r - k_2 y - k_3 \dot{y} \qquad (17)$$

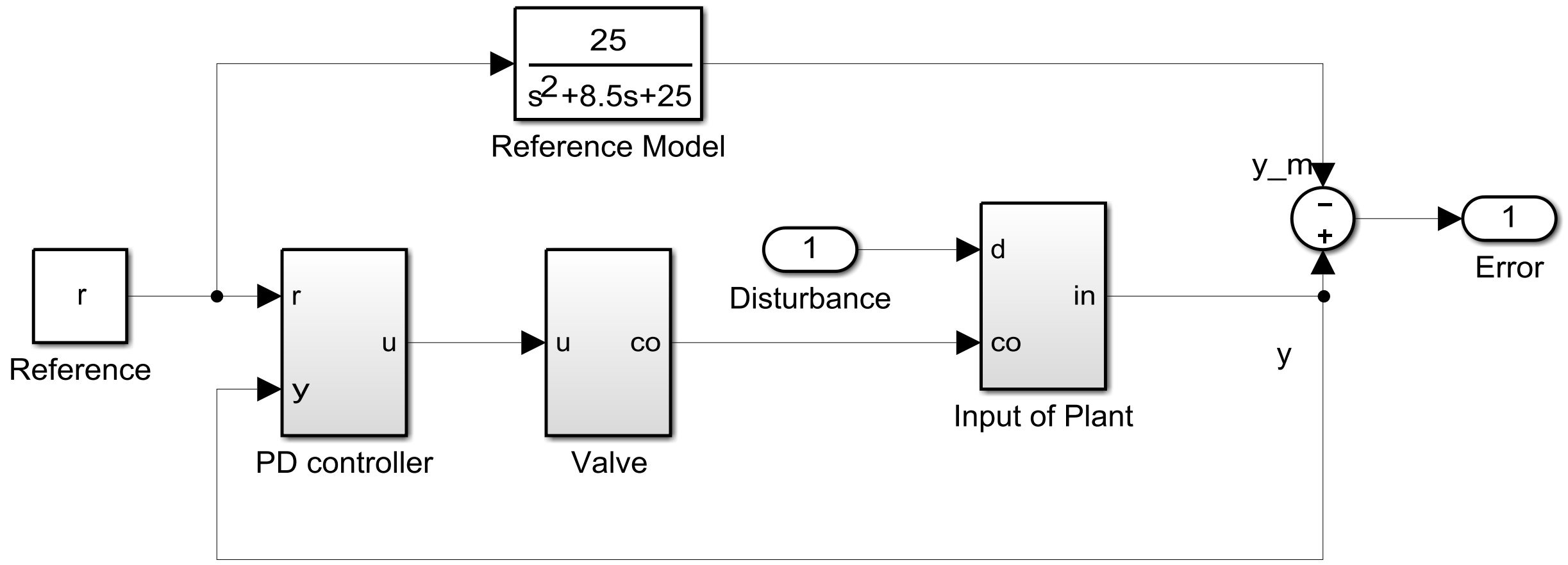

**Fig. 9. Schematic of adaptive controller (MRAS)**

Where r is the reference signal and y is the output of the plant. Based on MIT rule, controller parameters ($\theta$) must be selected in order to minimize a performance measure

$$J = \frac{1}{2} e^2 \tag{18}$$

Where e represents the error between the output of the plant ( $y$ ) and the output of the reference model ( $y_m$ ). Therefore, the parameters are selected based on the following relation which is regarded as MIT rule:

$$\dot{\theta} = -\gamma \frac{\partial J}{\partial \theta} = -\gamma (y - y_m) \frac{\partial y}{\partial \theta} \tag{19}$$

Where $\gamma$ is adaptation gain. Valve is a nonlinear element. Variable x is introduced to simplify valve linear mode representation and it is the output of the valve in linear mode.

$$x = G_v(s) u \tag{20}$$

To consider the valve saturation effect, a new function is defined as

$$co = \begin{cases} 0.25 & x \ge 0.25 \\ x & 0.05 \le x \le 0.25 \\ 0.05 & x \le 0.05 \end{cases} \qquad (21)$$

Therefore, as input to the compressor we have

$$y = d + co \qquad (22)$$

Disturbance is independent of controller parameter θ because its source is out of the control loop. Then

$$\frac{\partial d}{\partial \theta} = 0 \quad \to \quad \frac{\partial y}{\partial \theta} = \frac{\partial d}{\partial \theta} + \frac{\partial co}{\partial \theta} = \frac{\partial co}{\partial \theta} \qquad (23)$$

$$If \quad x \ge 0.25,\ x \le 0.05 \quad \to \quad \frac{\partial y}{\partial k_1} = 0 \qquad (24.a)$$

$$If \quad 0.05 \le x \le 0.25 \quad \to \quad \frac{\partial y}{\partial k_1} = \frac{\partial x}{\partial k_1} = \frac{r}{2s+1} \qquad (24.b)$$

$$If \quad x \ge 0.25,\ x \le 0.05 \quad \to \quad \frac{\partial y}{\partial k_2} = 0 \qquad (25.a)$$

$$If \quad 0.05 \le x \le 0.25 \quad \to \quad \frac{\partial y}{\partial k_2} = \frac{\partial x}{\partial k_2} = \frac{-y}{2s+1} \qquad (25.b)$$

$$If \quad x \ge 0.25,\ x \le 0.05 \quad \to \quad \frac{\partial y}{\partial k_3} = 0 \qquad (26.a)$$

$$If \quad 0.05 \le x \le 0.25 \quad \to \quad \frac{\partial y}{\partial k_3} = \frac{\partial x}{\partial k_3} = \frac{-\dot{y}}{2s+1} \qquad (26.b)$$

PD controller should have a low steady state error for more efficiency. Considering requested speed, steady state error assumed to be zero and other features like optimization, speed, and small derivation gain with respect to proportional gain are selected according to what has been selected at section III. Besides, J has a nonlinear relation with controller parameter changes. It means that it is possible to find a local minimum instead of a global one. To overcome this issue and to provide suitable system performance, the designed controller ($k_1 = k_2 = 10, k_3 = 0.7$) would be acceptable for initialization of adaptive algorithm. Note that $k_1$ and $k_2$ are the same kind as $K_p$ and $k_3$ is the same kind as $K_d$ in PID controller structure.

### B. Simulation Results

Adaptive controller is evaluated in the presence of various disturbances (stable and unstable). Surge point (unstable) is 0.43 and desired point is 0.55. In the first simulation, system response to 0.35 mass flow rate disturbance with 1 second time constant is evaluated. Control system output and parameters are plotted in the following Figures. Simulation results, demonstrated in Fig. 10, shows that system is stable against surge smeared disturbance and operating point reaches desired point (most efficiency).

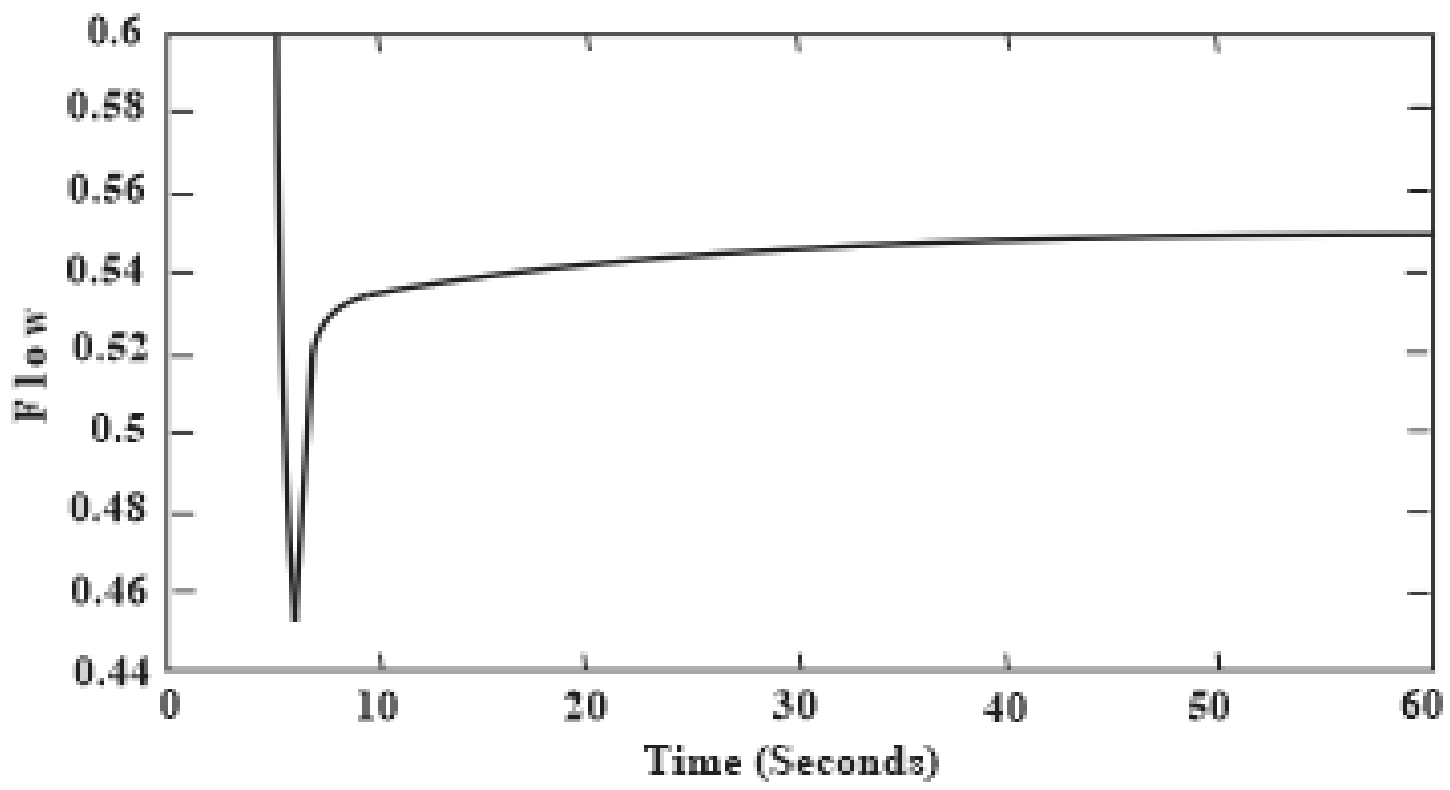

**Fig. 10. System output against disturbance 0.35**

$k_1$

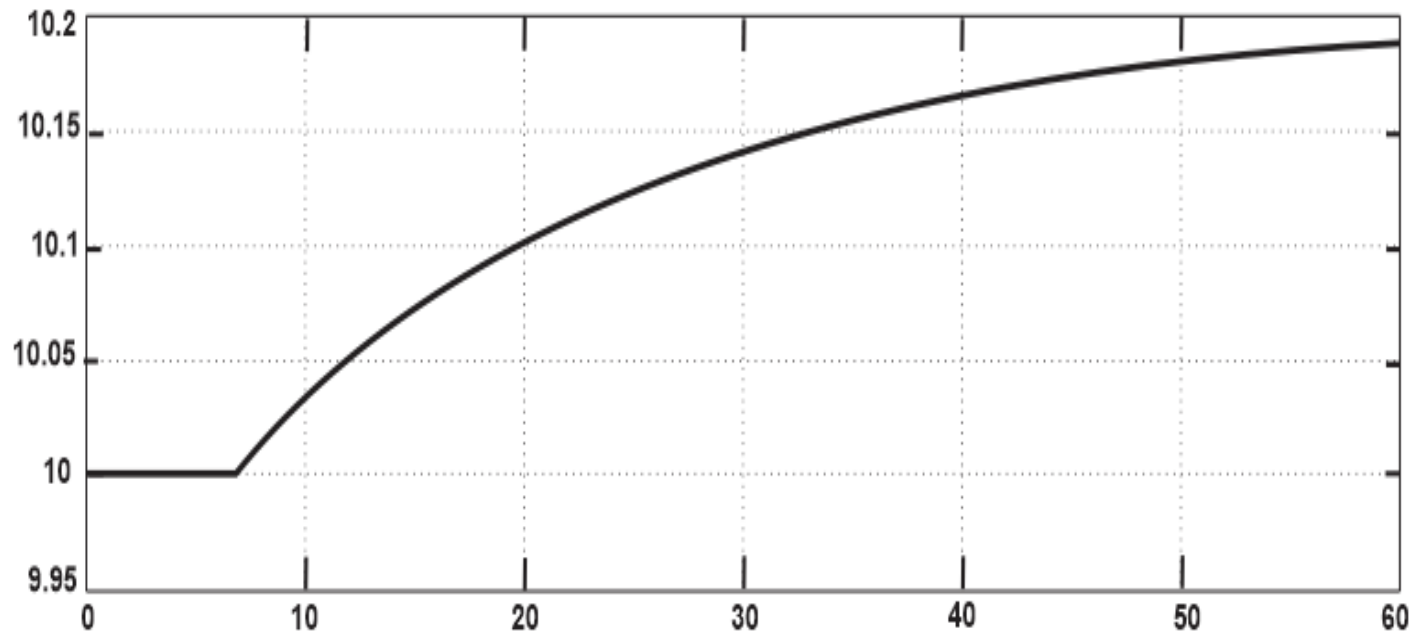

$k_2$

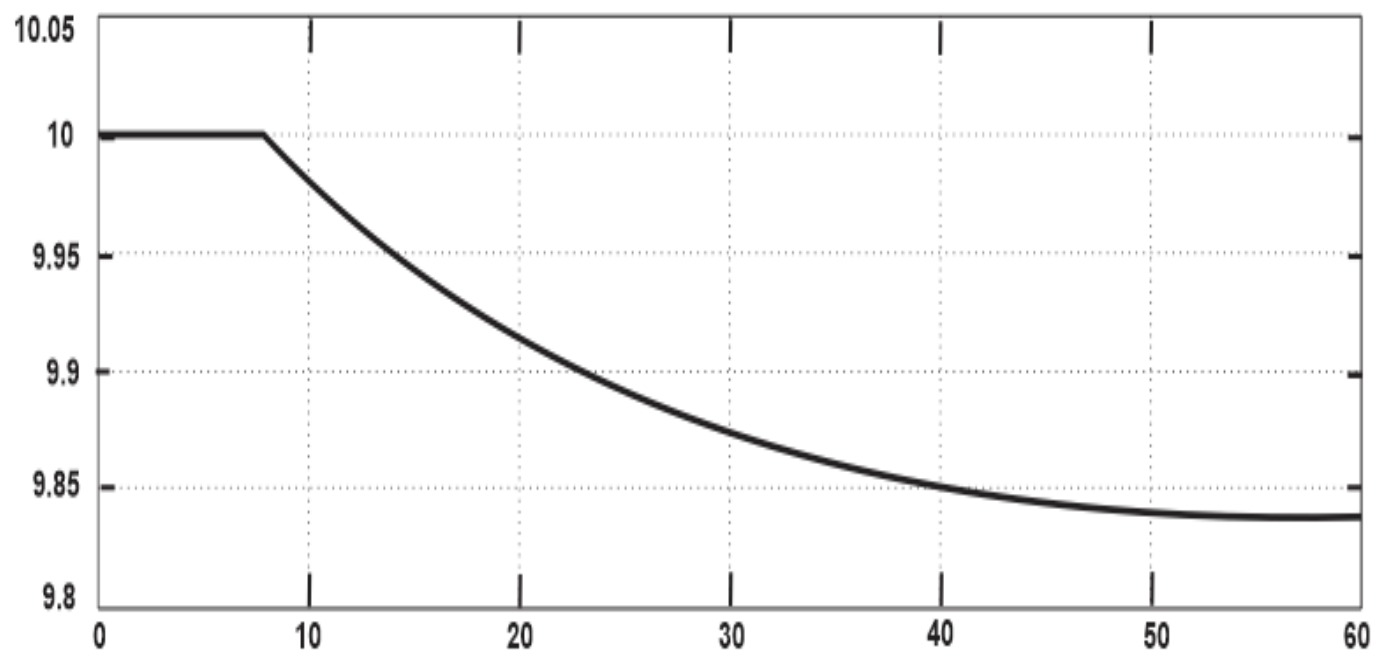

$k_3$

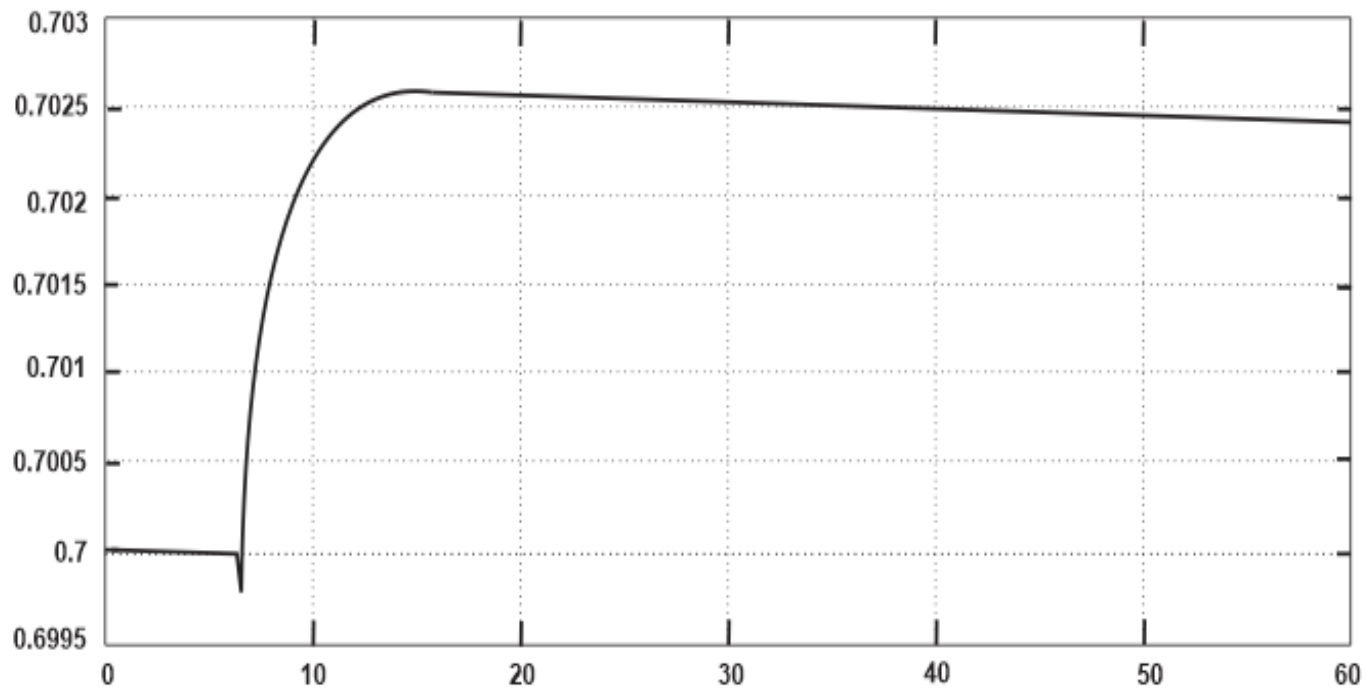

Time (Seconds)

**Fig. 11. Parameters variations with disturbance 0.35**

In the second simulation, system response to 0.45 mass flow rate disturbance with 1 second time constant is evaluated. Based on Fig. 12, controller reaches operation point to desired point. variation of parameters with 0.45 disturbance (Fig. 13) is less than that of 0.35 disturbance (Fig. 11); that is because deviation from desired point (0.55) with 0.45 disturbance is lower.

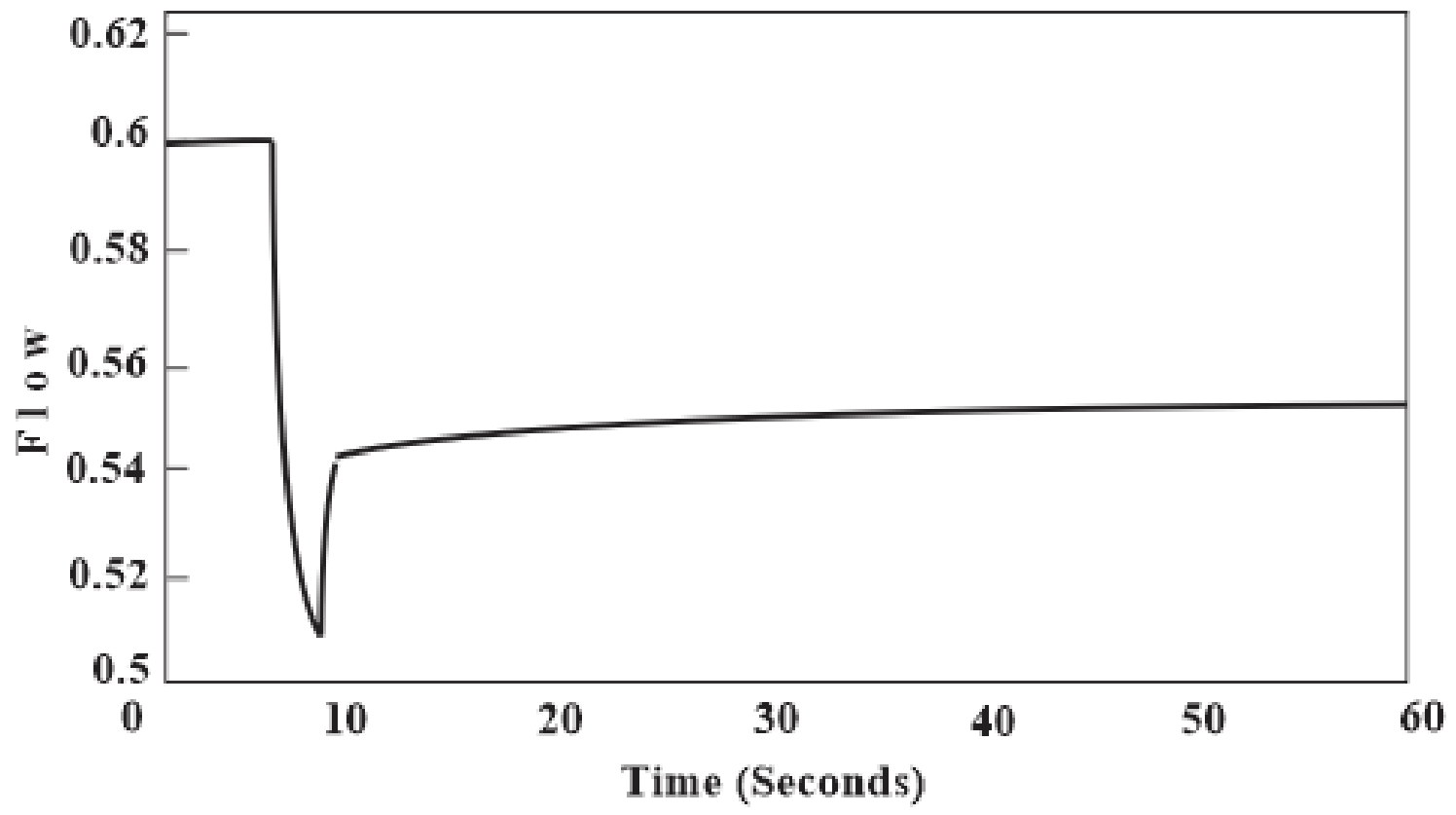

**Fig. 12. System output against disturbance 0.45**

$k_1$

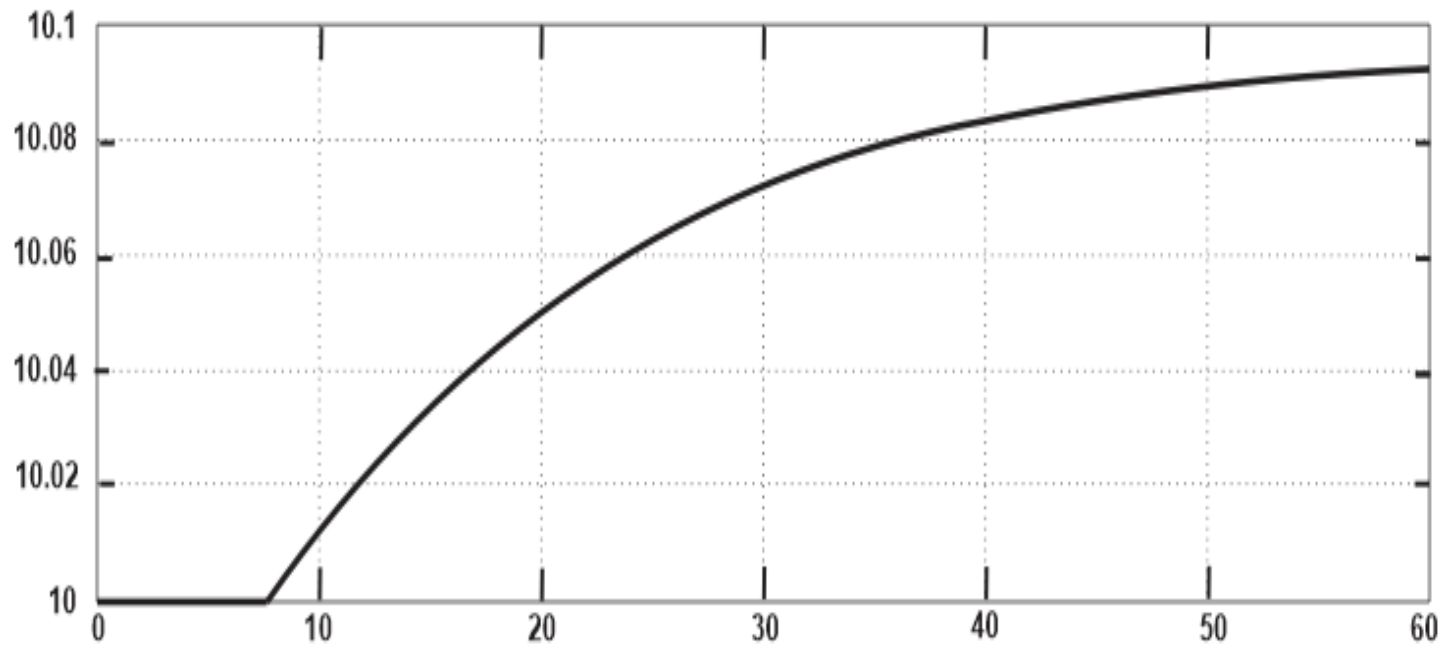

$k_2$

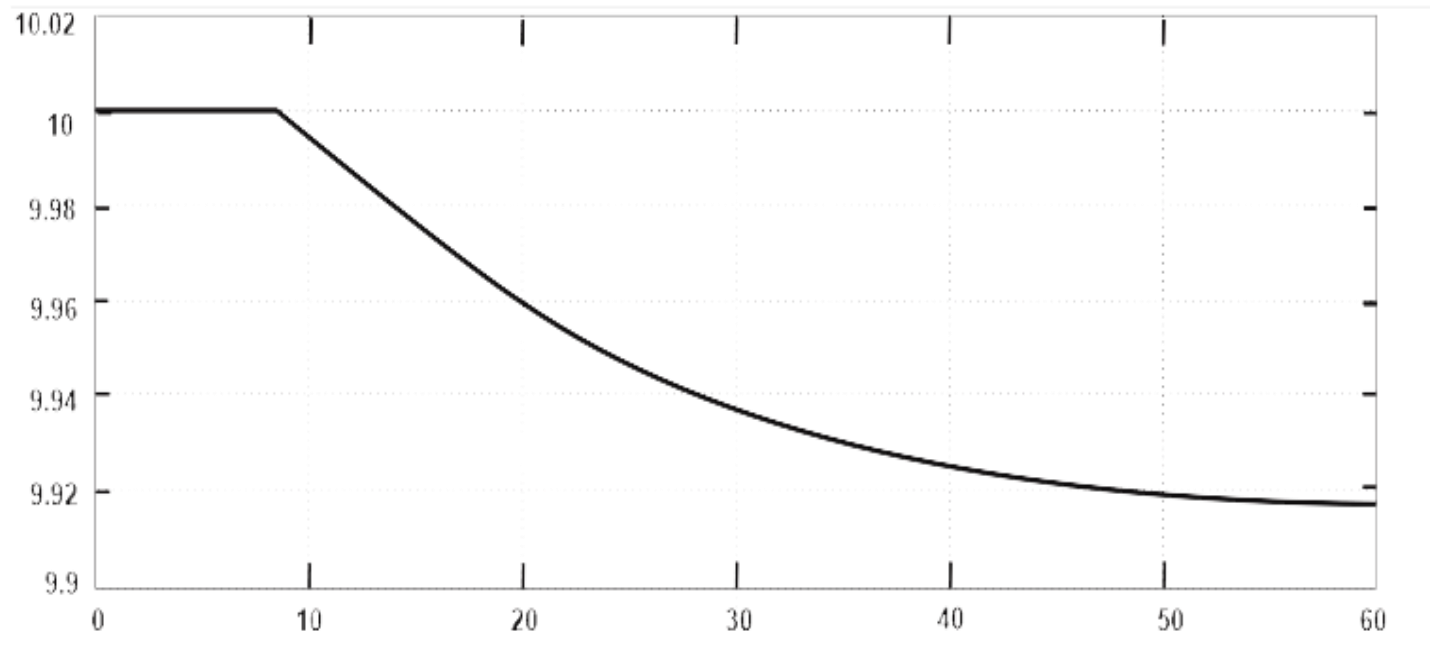

$k_3$

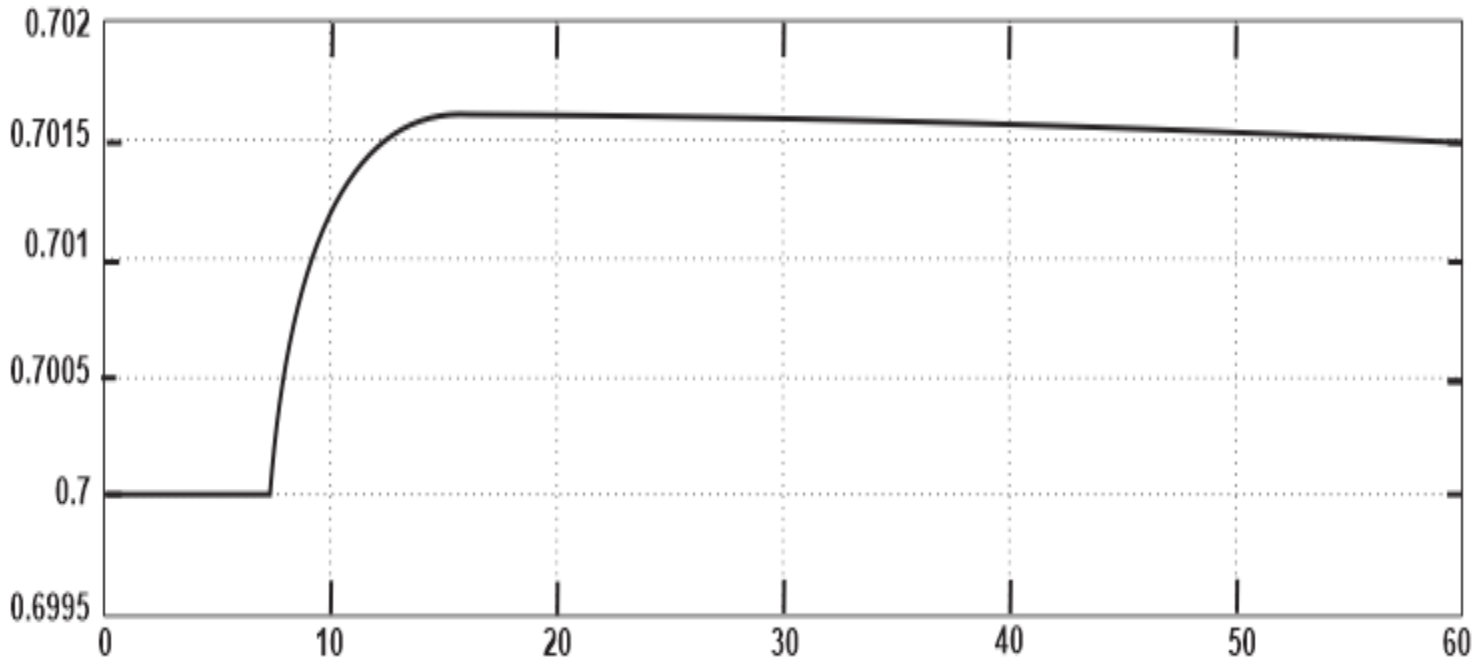

Time (Second)

**Fig. 13. Parameters variations with disturbance 0.45**

In third simulation, 0.6 disturbance with 1 second time constant enters to the system. Based on Fig. 14, controller cannot reach operation point to desired point; that is because the actuator is limited.

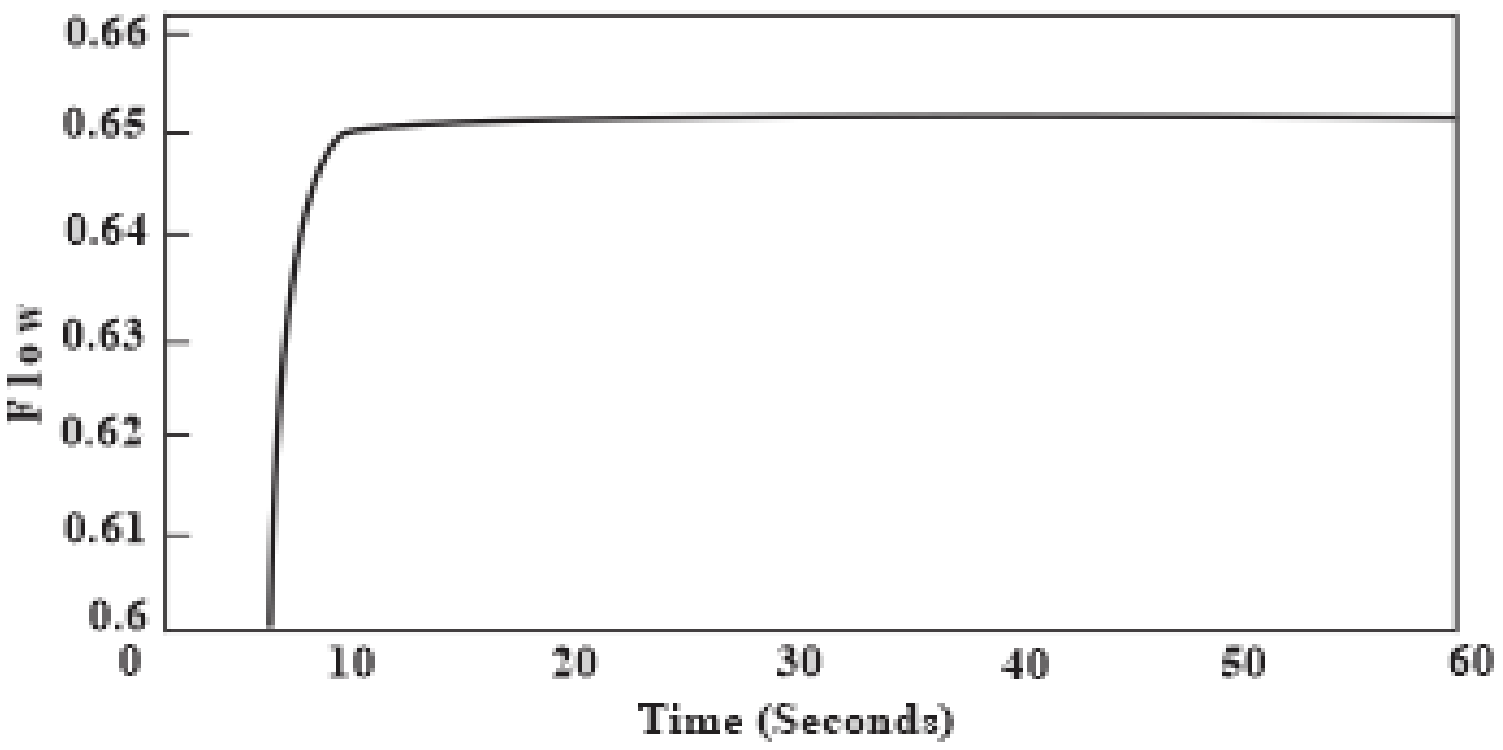

**Fig. 14. System output against disturbance 0.6**

Depicted results in Fig. 14 indicates that parameters don’t change. That’s because system output value is more than reference value, controller output is negative. Since actuator output is between 0.05 and 0.25, hence actuator output takes the constant value of 0.05 and adaptation doesn’t occur. In this simulation we get $k_1 = k_2 = 10, k_3 = 0.7$ and they haven’t changed at all.

### C. Stability Investigation (averaging method)[19]

Since based on MIT rule, parameters relation is

$$\dot{\theta} = -\gamma \frac{\partial J}{\partial \theta} = -\gamma (y - y_m) \frac{\partial y}{\partial \theta} \qquad (27)$$

then parameter stability results in stability or boundedness of error ( $y - y_m$ ). Based on the stability of the reference model ( $y_m$ ), output of control loop is also stable. Controller output is stable and bounded; therefore, stability of the plant and the parameters leads to the stability and boundedness of the output of the system. Parameters stability is investigated according to the averaging method. Based on equations (24.a)-(26.a), parameters of the system are adapted in linear actuator mode and actuator saturation mode as follows.

- Linear actuator mode:

$$\dot{k}_1 = -\gamma(y - y_m)\frac{r}{2s+1} \tag{28.a}$$

$$\dot{k}_2 = -\gamma(y - y_m)\frac{-y}{2s+1} \tag{28.b}$$

$$\dot{k}_3 = -\gamma(y - y_m)\frac{-\dot{y}}{2s+1} \tag{28.c}$$

- Actuator saturation mode:

$$\dot{k}_1 = 0 \tag{29.a}$$

$$\dot{k}_2 = 0 \tag{29.b}$$

$$\dot{k}_3 = 0 \tag{29.c}$$

In this case, parameter’s dynamics around operating point becomes a function of $\omega$ (note that $s = j\omega$), and the stability is proven with linearization of this dynamics and finding eigenvalues of the Jacobian matrix. The system is analyzed at $\omega = 0$, because controller reference or control loop input is a regular (constant) value Therefore, averaging method becomes simple such that “s” is replaced with zero in dynamic parameters equations. Hereafter parameters dynamic is a nonlinear function of the parameters and Jacobian matrix is yielded. In this Jacobian matrix, operating point of the parameters haven’t inserted until matrix is yielded based on the parameters and limits of parameters is obtained in analyzing stability.

- Linear actuator mode:

$$\left.\begin{aligned} u &= k_1 r - k_2 y - k_3 s y \\ y &= \frac{u}{2s+1} \end{aligned}\right\} \quad \rightarrow \quad y = \frac{k_1}{2s+1+k_2+k_3 s} r \tag{30}$$

$$\left.\begin{aligned} G_m &= \frac{25}{s^2+8.5s+25} \\ y_m &= G_m r \end{aligned}\right\} \quad \rightarrow \quad y_m = \frac{25}{s^2+8.5s+25} r \tag{31}$$

$$\dot{k}_1 = -\gamma (y - y_m) \frac{r}{2s+1} \quad \xrightarrow{s=0} \quad \dot{\bar{k}}_1 = -\gamma r^2 \left( \frac{k_1}{1+k_2} - 1 \right) \tag{32}$$

$$\dot{k}_2 = -\gamma (y - y_m) \frac{-y}{2s+1} \quad \xrightarrow{s=0} \quad \dot{\bar{k}}_2 = \gamma r^2 \left( \frac{k_1}{1+k_2} - 1 \right)\left( \frac{k_1}{1+k_2} \right) \tag{33}$$

$$\dot{k}_3 = -\gamma (y - y_m) \frac{-sy}{2s+1} \quad \xrightarrow{s=0} \quad \dot{\bar{k}}_3 = 0 \tag{34}$$

Here, Jacobian matrix and its eigenvalues are derived as:

$$J = \gamma r^2 \begin{bmatrix} \dfrac{-1}{1+k_2} & \dfrac{k_1}{(1+k_2)^2} & 0 \\ \dfrac{2k_1}{(1+k_2)^2} - \dfrac{1}{1+k_2} & \dfrac{k_1}{(1+k_2)^2} - \dfrac{2k_1^{\,2} * (1+k_2)}{(1+k_2)^4} & 0 \\ 0 & 0 & 0 \end{bmatrix} \tag{35}$$

$$eig(J) = \left( 0,\ 0,\ -\frac{(k_2 - \frac{1}{5k_1} + 1)^2 + \frac{1}{75} k_1^{\,2}}{(1+k_2)^3} \right) \gamma r^2 \tag{36}$$

Since parameters ( $k_1, k_2, k_3$ ) are positive (controller coefficients are considered positive), eigenvalues for each operating point related to different parameter sets and various values of $k_1, k_2$ and $k_3$ become zero or negative. Therefore, dynamics of parameters adaptation is stable which makes the controller stable in linear actuator mode. Adaptation gain ($\gamma$) isn't effective in stability but the more value it has, the more negative eigenvalues become and hence, convergence speeds up.

In Actuator saturation mode, parameters variation are zero (Eq. (29.a)) and for every parameter, operation point and various values of $k_1, k_2$ and $k_3$, eigenvalues of the Jacobian matrix become zero. Adaption dynamic of parameters is stable in resulted controller as well as in actuator saturation mode. It means that adaptive controller is stable in all ranges of operation. System stability is guaranteed but because of actuator limitation the system output may not reach to the desired point.

## VI. SIMULATION OF NORMAL PID

In this part, normal PID is simulated based on Ziegler-Nichols (Z-N) method [21] and table1(mentioned in part C. PID controller). Coefficients of PID in this simulation is $K_p = 10, K_i = 24, K_d = 1.$ In following figure is showed that normal PID controller can't guarantee stability and input flow of compressor is less than 0.43(left side of surge line or unstable region) after entering unstable disturbance(system response to 0.35 mass flow rate disturbance with 1 second time constant is evaluated). This subject can prove better performance of adaptive PID controller(see Fig.10) in comparison with normal PID with adaption of coefficient when disturbance is changed. Moreover, existing integral will reduce reaction and speed of controller when it should supply better and suitable input to compressor.

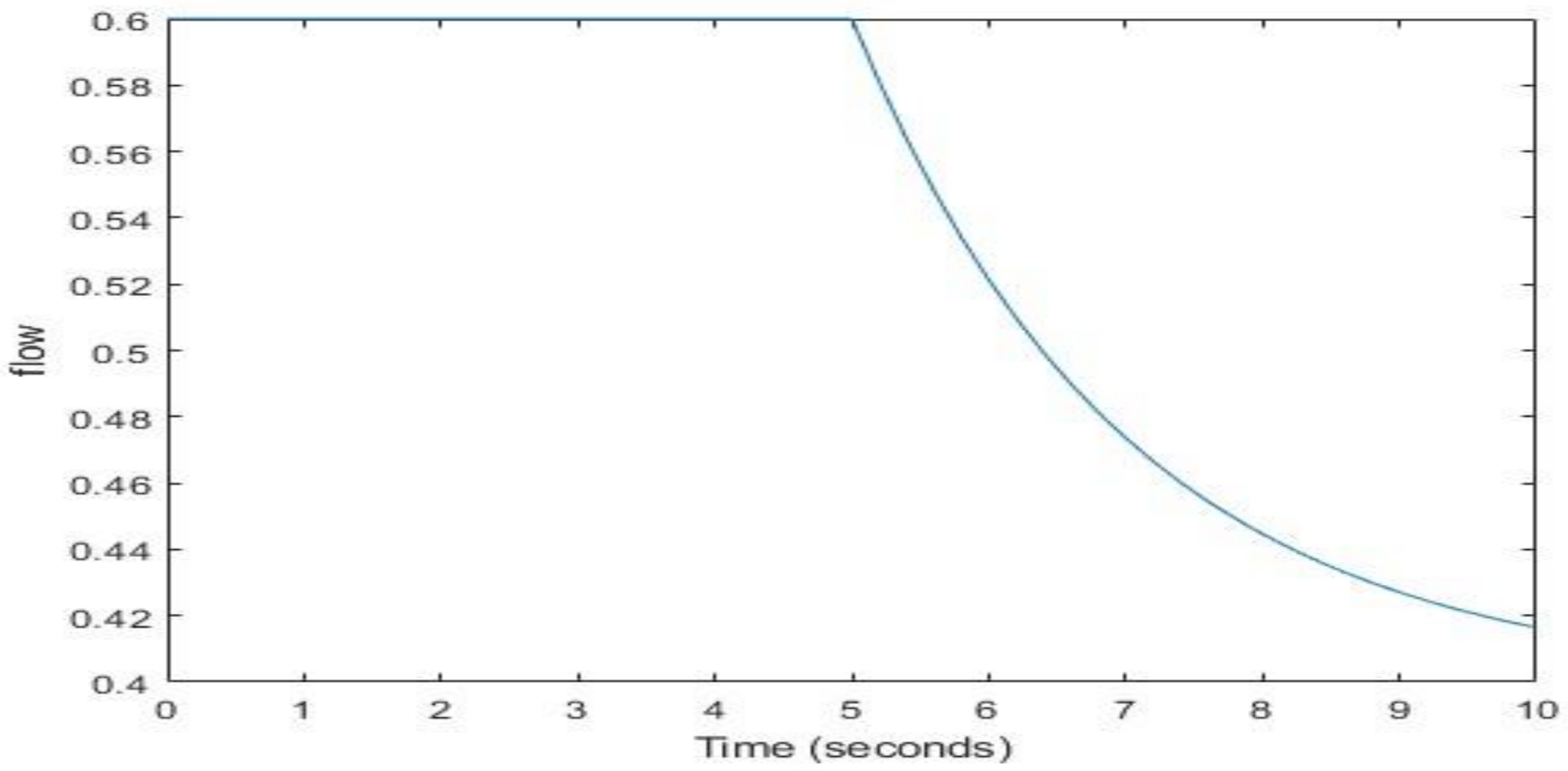

**Fig. 15. System output against disturbance 0.35 with PID**

## VII. Conclusion

In this paper, surge phenomena and compressor instability are analyzed. Based on nonlinear Greitzer state space equations, compressor is modeled. With Jacobian matrix and phase plane, stable and unstable regions of the system is extracted and existence of oscillations and limit cycle is proved. Validation data from previous study about especial compressor could clarify performance of mathematical method. Based on the equilibrium point analysis and compressor stability, it is concluded that stable and unstable regions can be determined with compressor input. Then based on an innovative approach, compressor model is deleted from control loop and its input is measured instead. Since upstream flow enters the system as a disturbance, it may change the operating point and hence it may reach to the unstable region. Designed PD controller is adapted based on the MRAS method. Adaptive PID controller could guarantee stability and tracking performance against disturbances merged surge. Proving stability of adaptive controller that is designed with gradient descend was investigated. Normal PID was designed for comparison with adaptive PID for evaluation that is better guaranteeing stability with adaption. Adaptation gain effects in convergence speed of controller parameters is demonstrated.

# I. Declarations

### A. Funding

Grant of FPI-UPC2021 .

### B. Conflicts of interest/Competing interests

The authors have no conflicts of interest to declare that are relevant to the content of this article.

### C. Code availability

All the simulations are available upon request.